\documentclass[9pt,twocolumn,twoside]{opticajnl}
\journal{opticajournal} 

\setboolean{shortarticle}{false}

\usepackage{lineno}

\title{Hyperspectral acquisition with ScanImage at the single pixel level: Application to time domain coherent Raman imaging}

\author[1]{Samuel Métais}
\author[1]{Sisira Suresh}
\author[1]{Paulo Diniz}
\author[1,2]{Siddarth Shivkumar}
\author[3]{Randy Bartels}
\author[4]{Nicolas Forget}
\author[1,*]{Hervé Rigneault}

\affil[1]{Aix Marseille Univ, CNRS, Centrale Med, Institut Fresnel, Marseille, France.}
\affil[2]{Department of Physics, University of Ottawa, Ottawa, Ontario K1N6N5, Canada}
\affil[3]{Morgridge Institute for Research, Madison, WI, USA}
\affil[4]{Cote d’Azur Univ, CNRS, Institut de Physique de Nice (INPHYNI), Nice, France}
\affil[*]{herve.rigneault@fresnel.fr}

\begin{abstract}
We present a comprehensive strategy and its practical implementation using the commercial ScanImage software platform to perform hyperspectral point scanning microscopy when a fast time dependant signal varies at each pixel level. In the proposed acquisition scheme the scan along the X axis is slowed down while the data acquisition is maintained at high pace to enable the rapid acquisition of the time dependant signal at each pixel level. The ScanImage generated raw 2D images have a very asymmetric aspect ratio between X and Y, the X axis encoding both for space and time acquisition. The results are X axis macro-pixel where the associated time depend signal is sampled therefore providing an hyperspectral information. We exemplified the proposed hyperspectral scheme in the context of time domain coherent Raman imaging where a pump pulse impulsively excites molecular vibrations that are subsequently probed by a time delayed probe pulse. In this case the time dependent signal is a fast acousto-optics delay line that can scan a delay of 4.5ps in 25$\mu$s, at each pixel level. We this acquisition scheme we demonstrate ultra-fast hyperspectral vibrational imaging in the low frequency range [10$cm^{-1}$, 150 $cm^{-1}$] over a 500 $\mu m$ field of view in 14ms (7 frames/s). The proposed acquisition scheme can be readily extended to other applications requiring to acquired a fast evolving signal at each pixel level.
\end{abstract}

\setboolean{displaycopyright}{false} 

\begin{document}

\maketitle

\section{Introduction}
Current research in advanced microscopy requires the implementation of powerful control software that can handle the control of a multitude of scientific instruments such as lasers, modulators, micro positioning translation and rotation devices, galvanometer scanners and detectors. They have also to integrate these scientific instruments into a hierarchical streamline of actions to perform specific tasks enabling basic operations such as imaging, z-stack, mosaicing but also more advanced functionalities such as frequency, polarization, delay sequences that can be simple scans or more advanced random access exploration. Because the number of possible combinations of these multi-spectral, multi-positioning, multi-polarization, multi-delay... is difficult to anticipate and very specific to any development, there is a need to develop custom software that are tailored and dedicated to each instruments. Several options are available for the developers ranging from basic languages such as C++ or Python\cite{noauthor_pythonorg_2024} to more developed languages such as LabView \cite{noauthor_labview_nodate} or MatLab \cite{noauthor_mathworks_nodate} that have the advantages to integrate multiple build-in functions and drivers that ease the control of opto-electronic devices and their integration in a comprehensive software. Although any development is possible with these tools there exist more advanced semi-commercial platforms that directly enable the user to acquire data from a proposed controlled software featuring a graphical user interface (GUI). This is for instance the case of the Micro-Manager \cite{noauthor_micro-manager_nodate} and ScanImage \cite{noauthor_scanimage_nodate} softwares that provide the user with a functional framework and GUI to rapidly control standard opto-electronics devices that are commonly used in microscopy. Whereas both can perform wide field or point scanning microscopy, ScanImage is more dedicated to point scanning microscopy and readily controls the galvanometer scanners and the detectors into independent threads that optimize the imaging speed and data acquisition rate. ScanImage is a MatLab based software that enables the user to add his own script to perform real-time data processing once the image is acquired. As an example, we have recently developed a stimulated Raman microscope that provide real time histology images based on the image of the $CH_{2}$ and the $CH_{3}$ chemical bonds present in the tissue sample\cite{appay_live_2023}. Using a Matlab script a virtual histology image could be generated in real time while exploring the sample [Movie at Ref to https://www.youtube.com/watch?v=R5mFdJBYpGE].
However, apart from simple Matlab add-on scripts that are easy to add, it is more challenging to enter into ScanImage intimate structure to change the scanning sequence organization. For instance the basic ScanImage operation mode is designed to acquire an analog signal at each spatial point in the scanned sample, therefore generating an image from the multitude of acquired signal at each sample positions. This is for instance what is commonly used to perform confocal fluorescence microscopy. ScanImage also allows for the synchronization of the acquisition data stream with an external clock to define an acquisition window relative of this external signal. When implemented with advanced high speed vDAQ  this allows fluorescence lifetime imaging (FLIM) where the analog signal from photon counting detector is rapidly digitized relative to the excitation laser pulse. However this FLIM modality is very specific and cannot be easily extended to other situations that necessitate to perform a multitude of time measurements for each spatial pixel. This is for instance the case in hyperspectral images, where at each pixel one needs to measure a spectrum.
In this work we are interested in using ScanImage to generate multidimensional images, that are images where the spatial (x,y) dimensions are complemented with other dimensions that can be a spectrum, a delay line, or any other time dependent signals that need to be recorded at each pixel. We consider that the time dependent signal follows a sequence and that this sequence needs to be fully acquired at one spatial (x,y) point before the system moves to the next point where the sequence is acquired again. As an example, let's consider the hyperspectral image case where at each pixel a sequence of frequencies illuminates sequentially the sample, the time dependant signal would be the transmission of the sample when it is illuminated by the various frequencies coming, for example, from a fast wavelength tunable lamp or laser. This is also what happens in pump-probe experiments on which we focus in this work. Here the time dependant signal is the transmission of a probe pulse that is time-delayed as compared to a pump pulse by a fast delay line. Here again, the full time trace evolution of the probe pulse transmission needs to be fully acquired at each spatial pixel (where the pump and the probe pulses are focused) before moving to the next pixel where the probe pulse transmission with varying delay needs to be acquired again.
Here, we address this problem with ScanImage and present a possible imaging scheme where the scanning beam is slowly moved along the x axis to enable for the fast delay line to scan. This enable to perform multiple time point measurements (each corresponding to a specific delay of the delay line) within a macro-pixel (defined as the distance x that is scanned during the time it takes for the delay line to perform its time scan) before moving to the next macro-pixel where the process is repeated. After a full x line acquisition, the scanner moves along y to the next line where the process is done again.

We believe they are multiple situations where this scheme and the associated ScanImage software could be beneficial to the users. This would be the situations where a relevant quantity moves quickly at each pixel while performing the image scan. This could be a polarization state, a frequency, a voltage, an intensity...  The scope of this paper is to describe the above mentioned scheme, to illustrate it in the case of time domain pump-probe coherent Raman imaging and make the software available to the community. 

The paper is organised as followed. In a first part we will briefly present the coherent Raman context and the associated implemented experiment, then we present the data acquisition scheme and illustrate the results. A supplementary information provides more technical details and makes available the developed ScanImage MatLab based software.

\begin{figure}[t]
    \centering
    \includegraphics[width=\linewidth]{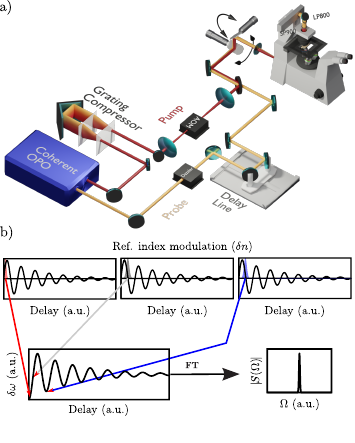}
    
    \caption{\label{fig:Opticalsetup3D} a) General sketch of the optical setup. b) Physical picture of the detected signal : the interaction of the probe pulse with the time modulation of the refractive index leads to shifts in its spectrum. A Fourier transform of the time signal  leads to a reconstructed spectrum.}
\end{figure}
\section{Time domain coherent Raman imaging: Acquiring a time dependent signal at each pixel}
Coherent Raman has been initially developed as a powerful spectroscopy technique to record the vibrational spectra of molecules when they are coherently driven by a laser field and subsequently probed by a second probe field \cite{levenson_coherent_1977,eesley_coherent_1981}. The technique has evolved in the early 2000's towards coherent Raman imaging where the vibrational signal from molecules was recorded at each spatial point of a sample to create an image \cite{cheng_coherent_2004,cheng_vibrational_2015,rigneault_tutorial_2018}. A variation of the technique follows an early scheme where the vibrational states of molecules are coherently excited by a single short pump pulse focused at a sample point. The molecular vibrations that are phase locked due to the short pump pulse impulsive excitation induce a refractive index transient variation that oscillate and decay in time. This transient refractive index is the direct signature of the molecular vibrations ringing and can be probed in time by a second probe pulse that is time delayed as compared to the initial short pump pulse \cite{ruhman_intramolecular_1987} (Fig. 1). The impulsive excitation enables to excite, and subsequently probe, all the molecular vibration frequencies that are lying in the pump pulse spectral bandwidth \cite{bartels_low_2021}. This scheme has been implemented recently in microscopy using focused scanning laser beams to generate images where the refractive index transient is recorded at each spatial position in a sample using the induced Kerr effect \cite{raanan_vibrational_2018} or the spectral shift \cite{raanan_impulsive_2018} experienced by the probe pulse. This later scheme takes advantage of the red or blue spectral shift experienced by the probe pulse if it interacts with a rising or decreasing refractive index, respectively (Fig.\ref{fig:Opticalsetup3D}). This spectral shift is detected as an intensity variation through a spectral filter (high or low pass) with its cut-off centered on the probe pulse spectral profile. The probe pulse spectral shift evolution over time provides a direct measure of the refractive index transient whose Fourier transform is the molecular vibrational spectrum (Fig.\ref{fig:Opticalsetup3D}).
We have recently improved the technique using an ultra-fast acousto-optics delay line (Dazzler, Fastlite) that can scan the probe pulse delay across 3.5ps in only 25$\mu$s\cite{raanan_sub-second_2019}. In our imaging scheme we use a pixel dwell time of 25$\mu$s that corresponds to one optical delay line scan and acquire during this time the refractive index transient through the probe spectral shift\cite{shivkumar_selective_2023}.\\
For the developer the problem breaks into (i) scanning the focused pump and probe laser beams across the sample and (ii) acquiring, at each pixel, the probe pulse spectral shift in 25$\mu$s. In this case the time dependent signal of the introduction section is the probe pulse spectral shift.

\begin{figure}[t!]
    \centering
    \includegraphics[width=\linewidth]{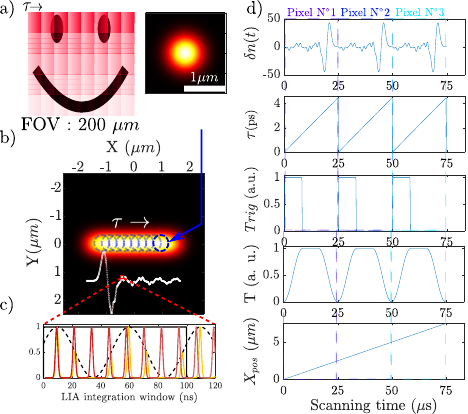}
   
    \caption{\label{Fig:scanning_scheme}a) Shematic description of the scanning process over a field of view containing a smiley face. b) The 'macro-pixel' PSF is defined by the red small bars where the real PSF (top insert) scans slowly the sample along the X axis during the time $\tau$. b) The 'macro-pixel' is therefore elongated along the X axis and has a dimension of $\sim$2$\mu$m while is dimension along the Y axis is $\sim$ 0.5$\mu$m. c) One time delay point corresponds to a LIA integration time (typically $\sim$300ns, here represented along 120ns), the pump pulse is represented in red while the time sweeping probe pulse in yellow, the AOM modulation is represented with the dashed line. d) Variations of the important parameters as a function of scanning time : from top to bottom, the apparent refractive index change $\delta(n)$ resulting from the vibrational coherent excitation, the pump/probe delay $\tau$ , the delay triggers $Trig$ (see Data Acquisition section), the transmission of the Dazzler $T$ and the position of the galvo-scanner $X_{pos}$.}
\end{figure}

\subsection{Optical Setup}
The optical setup is described in\cite{raanan_sub-second_2019} and the general layout is shown in Fig.\ref{fig:Opticalsetup3D}. Briefly, a fs OPO (Coherent) provides both the pump and the probe pulse trains (80MHz, 170fs pulse duration) . The probe is set at 800nm whereas the pump is set at 1045 nm. In order to get pulses that are close to Fourier-transform limited at the sample plane, the pump dispersion is compensated with a grating pair compressor (Fig.\ref{fig:Opticalsetup3D}). The pump pulse is then focused in an  AOM (AA-optoelectronic) for amplitude modulation at $f_{mod}=12$ MHz.
The probe pulse train is directed through an acousto-optics programmable dispersive filter (AOPDF) (Dazzler WB, Fastlite. Previous studies\cite{schubert_rapid-scan_2013,audier_pump-probe_2017} have shown that this AOPDF can operate as an ultra-fast delay line, enabling a swift 4.5 ps scan within just 25$\mu$s. This functionality is achieved using a birefringent crystal and an acoustic wave that diffracts each of the incoming vertically polarized probe pulses onto the horizontal crystal axis. As the acoustic wave takes 25$\mu$s to propagate through the crystal, each probe pulse's interaction with the acoustic wave determines its optical path and subsequent time delay, given the different refractive indices along the ordinary and extraordinary axes. Additionally, the Dazzler functions as a temporal pulse shaper, allowing the transfer of the temporally shaped acoustic wave onto the optical wave. In our work, we utilize this feature to negatively chirp the probe pulse, to compensate for the positive dispersion inherent in the setup.
Note that the probe pulse train goes through a mechanical delay line, this one remains static in an experiment, but it allows to adjust the 4.5ps time span of the fast Dazzler within the transient refractive index time trace.  
Finally, the pump and probe pulse are recombined with a dichroic mirror and send through the scanning mirrors of a custom made point scanning microscope.
The detection scheme looking for the probe pulse frequency shift versus its delay to the pump pulse arrival time will be described in the next section.

\section{Combining imaging and time spectroscopy with ScanImage}

\subsection{Hyperspectral imaging with ScanImage at the single pixel level}
At the first glance, our task is simple, it breaks into (i) placing the pump and probe beams at one location (x,y) on the sample (=a pixel) using the scanning mirrors, (ii) acquiring a time delay trace in 25$\mu$s (the time is takes to the Dazzler to scan 4.5ps) and to repeat this operation on the next pixel. However,in the commercial software ScanImage, the control of the scanning mirrors is limited to a two dimensional acquisition of data, one for each dimension X and Y. In order to perform imaging and time spectroscopy we have to acquire a data-cube (x, y and time delay), for that we have chosen a strategy that wraps the 'time' dimension into the X dimension as described below. In order to perform time spectroscopy at each pixel we slow down the scan along the X direction by a factor of several hundreds, but perform data acquisition at high speed while the fast Dazzler delay line is scanning. In other words, we move very slowly along the X axis to let the time for the time delay line to perform its scan. This leads to the construction of 2D images of $(N_x \times N_t) \times N_y$ which have 'macro-pixels' with a high aspect ratio. Indeed, while the scanning mirrors 'stays on a spatial x macro-pixel', the DAQ card acquires hundreds of sample in 25$\mu$s as the delay between the pump and probe is scanned. At the end of the X line, the scanning mirrors move in the Y direction and start to image the next X line in the same way. We have hence wrapped the time dimension in the X direction and left the Y one unaltered. This is represented in Fig. \ref{Fig:scanning_scheme} a). \\
The proposed scheme makes ScanImage compatible with hyperspectral imaging, at the single pixel level. It is also easily extendable to any other situations where a time dependant signal is rapidly acquired at each pixel during the scanning process.

\subsection{Data Acquisition}
\begin{figure}
    \centering
    \includegraphics[width=\linewidth]{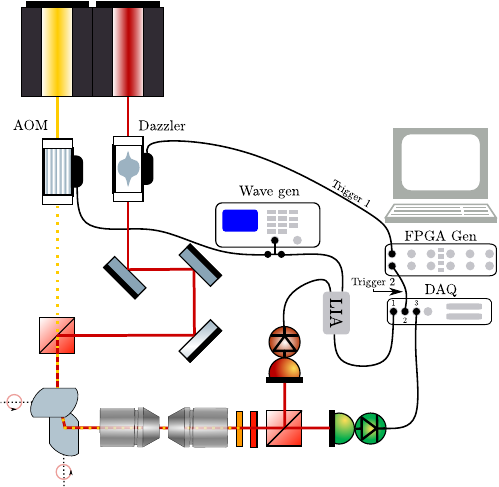}
   
    \caption{ \label{Fig:Setup electronic} Electronic layout of the experiment, two trigger signals generated by an FPGA synchronize the delay line (Dazzler) and the data acquisition (DAQ). The red photodiode corresponds to the Raman channel and the green one to the transmission channel.}
\end{figure}

The acquisition of the time-spectroscopy data requires the synchronization of the different electronics at play. This is represented in Fig.\ref{Fig:Setup electronic}. The first part is the lock-in-amplifier (LIA), whose reference signal is the 12MHz RF signal that modulates the AOM, this signal is generated by an external wave generator. Controlling the phase of the LIA allows to maximize the demodulated signal. Next, (i) the internal clock of the Dazzler RF generator (that launches the delay line) has to be synchronized with the internal clock of the DAQ. In order to accurately synchronized the Dazzler with the DAQ acquisition, for each time window where the delay line operates a scan, we use two identical digital trigger signals (up to 20MHz) generated from the FPGA that is available on the DAQ card's. These trigger signals repeat themselves after each delay scan (that is also the macro-pixel dwell time) (see panel 3 in Fig. \ref{Fig:scanning_scheme} d) . The first trigger it sent to the Dazzler to launch the Dazzler RF acoustic wave that will produce the delay scan. The second trigger, identical to the first, is fed to a DAQ fast acquisition channel as a synchro signal. With this it is possible to properly synchronize the triggering of the delay scan and the acquisition of the data.

The only parameter at play is the internal Dazzler RF delay between the acoustic wave trigger and the actual start of the delay scan, this is determined thanks to a calibration sample (a BGO crystal), and definitely fixed. 
Finally, the data acquired for an image corresponds to three 2D matrices of $(N_x \times N_t) \times N_y$, each corresponding to a DAQ analog acquisition channel. The DAQ channel 1 is the LIA data that corresponds to the Raman data ((see panel 1 in Fig. \ref{Fig:scanning_scheme} d); the DAQ channel 2 contains the Dazzler triggers for proper data synchronization (see panel 3 in Fig. \ref{Fig:scanning_scheme} d) and the DAQ channel 3 contains a signal coming from a photo-diode that records the transmission of the sample while scanning across the laser beams and that allows to re-normalize the Raman data according to time dependant transmission window of the Dazzler (see panel 4 in Fig. \ref{Fig:scanning_scheme} d). Figure \ref{Fig:scanning_scheme} provides a time domain representation of the various signals at play.

\subsection{Data Processing}
During the acquisition, there is a jitter that comes from the clock of the DAQ card, especially between each scanned line, as the scanning mirrors flight-back time is not necessarily a multiple of the DAQ time sample. The DAQ channel 2, which contains the trigger of the Dazzler scans, allows to properly register at each macro-pixel the Raman (DAQ channel 1) and the transmission (DAQ channel 3) data to the corresponding $(x,y,\tau)$ voxel. This allows to remap the two dimensional data to the corresponding hyperspectral cube, which is a 3D matrix of size  $(N_x,N_y,N_\tau)$. 
The final result is two $(N_x,N_y,N_\tau)$ 3D matrices which corresponds to the Raman signal ((Fig.\ref{Fig:Data representation} d) and the transmission (Fig.\ref{Fig:Data representation} e).\\
Most samples require more than a single 4.5ps Dazzler time window delay scan to properly determine the FWHM of their vibrational frequencies and we have implemented the possibility to record multiple Dazzler time windows shifted by the mechanical delay line. It is then necessary to recombine those different Dazzler delay windows to create a single time-spectroscopy signal, at each macro-pixel, over the total pump-probe delay investigated. Each time window is modulated across the delay by the transmission of the Dazzler. In order to properly stitch those windows we first normalize each delay graph by its corresponding transmission. This leads to an attenuation at the edges of the delay window. The recombination of the window is performed by interpolating data around the crossing of the falling transmission curve of window $n$ with the rising transmission of window $n+1$. As the expected signals are sums of decaying sinusoids, we use an interpolating algorithm that enforce continuity of the signal and its first derivative to minimize the error on the reconstructed spectrum. The data then consists of one 3D matrix of size $(N_x \times N_y \times (N_\tau \times N_{window})$. In order to obtain the hyperspectral cube, one then simply computes the fast Fourier transform of this time-spectroscopy signal for each spatial pixel.

\subsection{Vibrational resolution and Bandwidth}
 Limitations on the ability to detect a Raman line come from both the excitation and the read-out mechanism. 
 The first limitation is the bandwidth of the pump excitation laser. Indeed, with our pump bandwidth, one cannot excite Raman lines above 160 $cm^{-1}$. This limitation can be lifted with shorter laser pulses. Yet, the lock-in amplifier, with its intrinsic bandwidth, also limits our detection vibrational bandwidth, as faster vibrations will be averaged during its integration time. In order to compute this limitation one must take into account the time to delay scaling of the Dazzler, which is of 161 fs/us. The shortest integration time of our LIA is 100 ns which corresponds to a resolution in pump-probe delay of 17 fs. This corresponds to a frequency of 9.36THz which is 312$cm^{-1}$. This means that our detection electronic can not resolved vibrational frequencies that have a frequency above 312$cm^{-1}$. Any vibration faster than that time will hence be averaged out during the lock-in integration window. This limitation could be lifted with LIA with shorter integration time (with the drawback of a lower LIA filtering and lower detection efficiency).
 
\subsection{Spatial resolution}
 A feature of our detection scheme is its asymmetrical point spread function (PSF). Indeed, during the delay scan on each pixel, the scanning mirror keeps moving. This leads to an extended PSF in the X direction that we have named a 'macro-pixel' (see Fig. \ref{Fig:scanning_scheme}b). Along the Y axis the microscope resolution is around 1 µm (limited by the under-filling of the objective back focal plane). Along the X axis, the macro-pixel elongation $X_{Daz}$ is related to the distance scanned during the 25 µs of the Dazzer delay scan, which is related to the field of view. Indeed, the mirror scans continuously a region of size FOV in a time of $N_x \times \tau$. This means that during a single Dazzler window the mirror scans in the X direction a line of length $X_{Daz}=\frac{FOV}{N_x}$. This does not lead to a diminution in signal as both the pump and probe are scanned together. However this leads to an asymetric PSF whose aspect ratio is the FOV divided per the number of pixel $N_x$. This peculiar feature vanishes when the FOV is sampled at the limit of diffraction with one pixel per micron, as then the distance scanned by the mirror in 25$\mu s$ is one micron. During this scan, the DAQ card is sampling $N_\tau =270$ points which corresponds to the number of pump-probe delay samples that are acquired per pixel.


\section{Experimental Results}

\subsection{Raw Data}
\begin{figure}[ht]
    \centering
    \includegraphics[width=\linewidth]{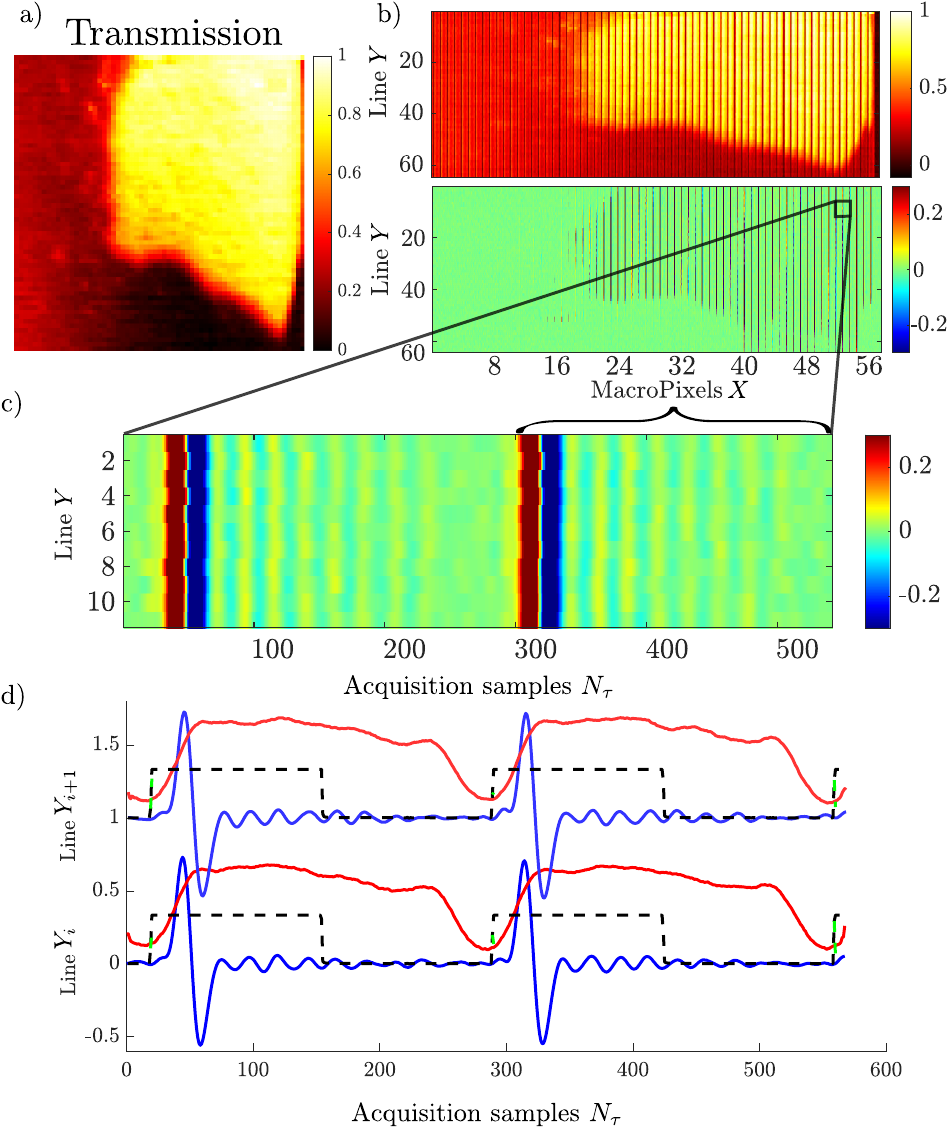}
    
    \caption{\label{Fig:Data representation}a) Transmission image of a BGO crystal. b) Raw data obtained from ScanImage for the transmission channel (top) and the Raman channel (bottom). c)  Zoom on two macropixels along the X direction for a 12 lines along Y showing the stong overlap between the pump and probe beams and the vibrational refractive index transient. d) Plots of the three channels, vibratinal refractive index transient (blue), Dazzler transmission (red) and the DAQ trigger (dashed line).}
    
\end{figure}

Our experimental setup is tested on a calibration crystal : a $5 \times 5$ mm piece of a germanate bismuth (BGO) crystal. This type of crystal exhibits a narrow vibrationnal resonance around 90$cm^{-1}$. Figure \ref{Fig:Data representation} a) shows the imaged corner of the BGO crystal as seen by the transmission photodiode (see Fig. \ref{Fig:Setup electronic} - green photodiode). This image has $(N_x.N_y)$=64*63 pixels ($N_x$ being the number of macro-pixel along X) and was acquired by summing the transmission channel over the course of the fast acousto-optic delay line). This averages the transmission over the course of the $N_\tau=270$ acquired sample points corresponding to a single Dazzler delay scan (see below). The Raman collected data follow the same acquisition scheme and consist of an image with a high aspect ratio of 270, which is equal to $N_\tau$, the number of acquired sample points in a single Dazzler scan window. The raw acquired transmission image is shown in the top panel of Figure \ref{Fig:Data representation} b) where the individual Dazzler transmission windows can be clearly seen along the X axis. The bottom panel of Figure \ref{Fig:Data representation} b) shows how the Raman data (i.e. the probe pulse frequency shift) develop in the image acquisition process. A zoom is shown in Figure \ref{Fig:Data representation} c) for 12 lines along the Y axis and 520 acquisition points along the X axis that correspond roughly to two Dazzler delay scans. A single Dazzler delay scan can be clearly identified from the contrasted sharp vertical red and blue lines that correspond to the temporal overlap between the pump and the probe, whereas the subsequent oscillations along the X direction are the coherent refractive index oscillations that develop within one delay scan. Figure \ref{Fig:Data representation} d) shows two intensity sections along the X axis for the refractive index variation (in blue) the Dazzler transmission (in red), and the DAQ trigger (dashed line). A Dazzler delay scan (25$\mu$s) corresponds to $N_\tau=270$ sample acquisitions to enable the capture of the transient refractive index, at each macro-pixel.  In this specific example the size of the macropixel along the X axis is $1 \mu m$. The repeatability from one Y line to another is correct but shows the necessity to normalize by the Dazzler transmission intensity (red line in Fig. \ref{Fig:Data representation} d)) for a proper acquisition of the refractive index time trace.

\begin{figure}[t!]
    \centering
    \includegraphics[width=\linewidth]{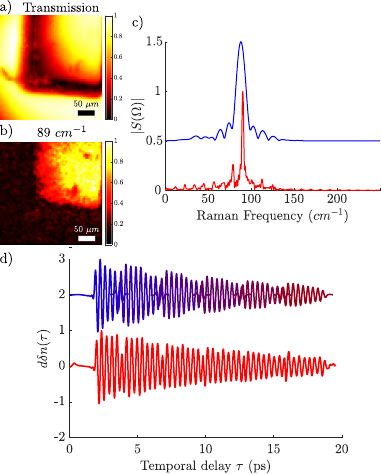}
    
    \caption{\label{Fig:bgo}a) Transmission image (300 $\mu m$ FOV) of the corner of a BGO crystal  b) Hyperspectral image obtained at 90 $cm^{-1}$ showing the chemical contrast c)  Spectrum obtained from a single Dazzler window (blue) and mutliple windows (red) demonstrating the spectral resolution. d) Time traces obtained for each position of the mechanical delay line (top, blue) and the interpolated time trace (red).}
    
\end{figure}

\subsection{Processed Data and Spectrum }
 Obtaining the 3D hyperspectral data cube ($x_{macropixel}$,y, Raman transient) requires to reshape the acquired 2D Data (Fig. \ref{Fig:Data representation}). For this we use the rising edges of the DAQ trigger channel which allows for a proper ordering and allocation of the data. This is why a proper synchronization of the the DAQ and Dazzler triggers is important. In practise we generate the Dazzler trigger before the DAQ trigger to take into account the Dazzler electronic internal delay, which is around few tens of microseconds. The last scanned line will contained truncated data if the synchronization is incorrect. Doing so we obtain two data cubes: one for the Raman channel ($x_{macropixel}$,y, Raman transient) and one for the transmission channel ($x_{macropixel}$,y,Dazzler transmission). As we have been using the spectral shift of the probe beam as the Raman transient\cite{raanan_impulsive_2018}, we obtain time traces proportional to the first derivative of the refractive index transient modulation induced by the pump pulse.
 The time-spectroscopy data is then processed in the following way for each pixel : the time dependant transmission is used to normalize the Raman time trace; we then proceed to filter out the overlap between the pump and the probe which yields a strong signal (see Fig. \ref{Fig:Data representation} c and d) using a 'tukey' window with a strong slope. This overlap strong signal would largely impact the spectrum. From this point, the basic fast Fourier transform (FFT) Matlab function is used to a zero-padded signal to yield the expected vibrational spectrum. This spectrum is shown in Fig. \ref{Fig:bgo} c, with the corresponding hyperspectral image at the 90$cm^{-1}$ peak (Fig. \ref{Fig:bgo} b). Generating an hyperspectral image over $64 \times  64$ $ \mu m$ with a $1$ $\mu m$ resolution  as shown in Fig \ref{Fig:bgo} c takes 130 ms ($\approx$ 8 Hz). When the FFT is performed with the data collected within single scan of the Dazzler window (that corresponds to 4.5ps) the spectral resolution is limited as shown in Fig.\ref{Fig:bgo} c (blue curve). Moving the mechanical delay line by steps of 3 ps, it is possible to collect multiple Dazzler delay windows that are used to build a longer time trace (Fig.\ref{Fig:bgo} d (blue curve)). This is done by interpolating the time traces at the points they overlap (Fig.\ref{Fig:bgo} d (red curve)). The FFT performed over a total window of about 18 ps leads now to a spectrum with increased spectral resolution (2 $cm^{-1}$ FWHM) (Fig.\ref{Fig:bgo} c (red curve)) . However, this multi time windowing slows down the imaging process, as moving the mechanical delay line takes 0.15 s, which means that this increased spectral resolution scheme lowers the frame rate to 0.3 Hz. 

\begin{figure}[t!]
    \centering
    \includegraphics[width=\linewidth]{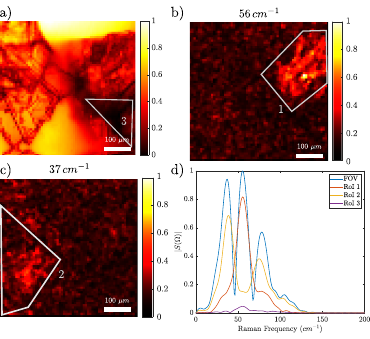}
    
    \caption{\label{Fig:aceto}a) Transmission image (500 $\mu m$ FOV) of a sample composed of a mixture of acetominophene and anthracene crystals; b) Hyperspectral image obtained at 56 $cm^{-1}$ highlighting acetominophene. c)  Hyperspectral image obtained at 37 $cm^{-1}$ highliting anthracene. d) Spectra obtained while averaging on either the whole FOV (blue), on Fig. \ref{Fig:aceto}b ROI 1 (red - acetominophene), on Fig. \ref{Fig:aceto}c ROI 2 (yellow - anthracene) and Fig. \ref{Fig:aceto}a ROI 3 (purple - void). Resolution 8 $\mu m$ along the X axis and 1 $\mu m$ along the Y axis. Hyperspectral acquisition speed: 14ms (7Hz frame rate). Single Dazzler delay scan of 4.5ps (spectral resolution: 10 $cm^{-1}$. }
    
\end{figure}
\subsection{Hyperspectral Imaging : Chemical detection}
To further demonstrate the capacities of the time domain coherent Raman imaging setup we have made a sample combining a crystal of Anthracene ($C_{14}H_{10}$) with smaller pieces of crystallized Acetominophene ($C_{8}H_{9}NO_{2}$). Both crystals are transparent and cannot be distinguished by the transmission image (see Fig.\ref{Fig:aceto} a)). The Raman spectrum averaged over the whole field of view shows three peaks, which corresponds to the contribution of both species (Fig. \ref{Fig:aceto} d) blue spectrum). When averaging the spectrum over specific region of interests (ROI) the resulting spectrum show the chemical distinction between the two species. For instance, Fig. \ref{Fig:aceto} b) shows the 56 $cm^{-1}$ image corresponding to Acetominophene whereas Fig. \ref{Fig:aceto} c) shows the 37 $cm^{-1}$ image corresponding to Anthracene. These hyperspectral images were processed with a Gaussian filter of size $2\times2$. The integrated spectra of ROI 1 on Fig. \ref{Fig:aceto} b) and ROI 2 on Fig. \ref{Fig:aceto} c) are shown in Fig. \ref{Fig:aceto} d) in red and yellow, respectively. They show the distinctive Raman peaks of Acetominophene  and Anthracene in this low frequency wavenumber region. Interestingly, the integrated spectrum of ROI 3 on Fig. \ref{Fig:aceto} a) show no distinctive Raman signature. The $64\times 63$ pixels hyperspectral coherent Raman images shown in Fig. \ref{Fig:aceto} were acquired at a rate of $\approx$ 7 frames/second (7 Hz) over a FoV of $500 \times 500$ $\mu m$. The resolution was 8 $\mu m$ along the X axis and 1 $\mu m$ along the Y axis. Here a single Dazzler time scan window was used.  These results demonstrate close to video rate chemical imaging in the vibration low frequency domain. Figure \ref{Fig:aceto} shows some outputs of the specific MatLab based user interface that we have developed that processes the hyperspectral Raman and transmission data cubes and outputs the hyperspectral images at selected frequencies, the spectrum of selected ROI and the integrated Raman spectrum images in real time. This is tool makes low frequency Raman imaging practical for investigations in pharmaceutical, material and possibly biology fields.

\section{Conclusion}
In this work we have developed a generic strategy that extends the ability of the ScanImage software to acquire hyperspectral images at the single pixel level. Our scheme allows to acquire a rapid time dependant signal at each pixel during the 2D (X,Y) spatial scan. We have exemplified our developed strategy and software in the case of time domain fast coherent Raman imaging where the time dependant signal is a delay line that rapidly scans (in 25$\mu$s) a 4.5ps time window for each pixel in the image. We have shown that this allows to acquire hyperspectral coherent Raman images over 500 $\mu m$ field of view in 14ms (7 frames/s). This outperforms by a factor of three our previous time domain coherent Raman acquisition speed\cite{raanan_sub-second_2019,shivkumar_selective_2023} and provides a user-friendly graphical user interface that has been developed as an extra layer on the commercial ScanImage software. Although our implementation was in the case of a fast delay line scan at each pixel, the proposed scheme is more general and can be readily applied with any time depend signal that need to be acquire at each pixel level. For instance, a fast rotating polarization\cite{hofer_high-speed_2017}, a time domain patterned illumination\cite{Scotte:19,heuke_spatial_2020} or any type of frequency encoding\cite{jiang_fast_2022,martins_openspyrit_2023} in single pixel detection scheme, to cite a few. Combined with real time decision and control taken during the scan time\cite{abouakil_adaptive_2021}the developed approach is expected to significantly extend the tool box of the instrument and software developers to advance the field of point scanning microscopy.

\section{Acknowledgment}
We acknowledge financial support from the Centre National de la Recherche Scientifique (CNRS), A*Midex (ANR-11-IDEX-0001-02), ANR grants (ANR-10-INSB-04-01, ANR-11-INSB-0006, ANR-16-CONV-0001, ANR-21-ESRS-0002 IDEC). This has received funding from European Union’s Horizon 2020 (EU ICT 101016923 CRIMSON and EU EIC 101099058 VIRUSONG) and European Research Council (ERC, SpeckleCARS, 101052911).
\bibliography{biblio}



\end{document}